\documentclass[journal = aamick,layout=twocolumn]{achemso}

\pdfoutput=1
\usepackage[fontsize=10pt]{scrextend}
\usepackage{stfloats}
\usepackage{eurosym}

\usepackage[utf8x]{inputenc}
\usepackage{amsmath}
\usepackage{amsfonts}
\usepackage{amssymb}
\usepackage{graphicx}
\usepackage{natbib}
\usepackage{hyperref}

\author{Stefan J. Eder}
\email{stefan.eder@ac2t.at}
\phone{+43 (0)2622 81600 161}
\fax{+43 (0)2622 81600 99}
\affiliation[AC2T research GmbH]
{AC2T research GmbH, Viktor-Kaplan-Stra\ss e 2, 2700 Wiener Neustadt, Austria}

\author{Ulrike Cihak-Bayr}
\affiliation[AC2T research GmbH]
{AC2T research GmbH, Viktor-Kaplan-Stra\ss e 2, 2700 Wiener Neustadt, Austria}

\author{Davide Bianchi}
\affiliation[AC2T research GmbH]
{AC2T research GmbH, Viktor-Kaplan-Stra\ss e 2, 2700 Wiener Neustadt, Austria}

\author{Gregor Feldbauer}
\affiliation[Hamburg University of Technology]
{Institute of Advanced Ceramics, Hamburg University of Technology, Denickestra\ss e 15/K, 21073 Hamburg, Germany}

\author{Gerhard Betz}
\affiliation[Vienna University of Technology]
{Institute of Applied Physics, Vienna University of Technology, Wiedner Hauptstra\ss e 8--10/134, 1040 Vienna, Austria}

\title{Thermostat Influence on the Structural Development and Material Removal during Abrasion of Nanocrystalline Ferrite}


\begin{document}

\begin{abstract}
We consider a nanomachining process of hard, abrasive particles grinding on the rough surface of a polycrystalline ferritic work piece. Using extensive large-scale molecular dynamics (MD) simulations, we show that the mode of thermostatting, i.e., the way that the heat generated through deformation and friction is removed from the system, has crucial impact on tribological and materials related phenomena. By adopting an electron-phonon coupling approach to parametrize the thermostat of the system, thus including the electronic contribution to the thermal conductivity of iron, we can reproduce the experimentally measured values that yield realistic temperature gradients in the work piece. We compare these results to those obtained by assuming the two extreme cases of only phononic heat conduction and instantaneous removal of the heat generated in the machining interface. Our discussion of the differences between these three cases reveals that although the average shear stress is virtually temperature independent up to a normal pressure of approximately 1~GPa, the grain and chip morphology as well as most relevant quantities depend heavily on the mode of thermostatting beyond a normal pressure of 0.4~GPa. These pronounced differences can be explained by the thermally activated processes that guide the reaction of the Fe lattice to the external mechanical and thermal loads caused by nanomachining.
\end{abstract}

\vspace{1cm}
{\bf Note:} This document is the unedited Author's version of a Submitted Work that was subsequently accepted for publication in \emph{ACS Applied Materials \& Interfaces}, copyright \textcopyright American Chemical Society after peer review. To access the final edited and published work see \url{http://pubs.acs.org/doi/full/10.1021/acsami.7b01237}.

Please cite this work as: 

S.~J.~Eder, U.~Cihak-Bayr, D.~Bianchi, G.~Feldbauer, and G.~Betz, ``Thermostat Influence on the Structural Development and Material Removal during Abrasion of Nanocrystalline Ferrite,'' {\em ACS Applied Materials \& Interfaces}, DOI: 10.1021/acsami.7b01237, 2017.


\section{Introduction}
\label{sect:Intro} 

The response of the near-surface microstructure of a work piece or a system component subjected to abrasive conditions can strongly influence its properties and behavior. This aspect of surface modification is of high interest to the surface finishing industry, where material is deliberately removed from the surface to obtain a desired roughness or texture, but also for tribological applications, where the -- usually unintended -- wearing of surfaces affects a system's performance and service life. Depending on the relative velocities of the surfaces in mechanical contact, the friction and possible plastic deformation occurring in the interface can lead to the generation of considerable amounts of heat, which has a large impact on the development of the microstructure close to that interface.  

The geometrical tolerances for components in various applications have increasingly entered into the nanometric length scale, e.g., in high-gloss finishing for optical purposes~\cite{Rebeggiani2013} or in nano/micro-electromechanical systems (NEMS/MEMS)~\cite{Rogers2016}. As the mechanisms governing material removal and wear, but also those responsible for changes to the crystal structure have their foundations in the atomistic nature of the component or work piece, it makes sense to consider the respective processes from a nanoscopic point of view. 

The shift in tribology towards the nano length scale took place over the last decades and showed that commonly used macroscopic tribological models as well as laws cannot be readily applied at this scale~\cite{Gotsmann2008,Jacobs2013,Wolloch2014}. Particularly, nanoscale wear processes~\cite{Bhushan1995,Gnecco2002,Hu2015} are not fully understood because of their highly complex nature~\cite{Kim2012}. Over the last years a high effort has been put into a better understanding of single-asperity contacts, which can be probed experimentally in detail using an atomic force microscope (AFM)~\cite{Szlufarska2008}. This approach is of high importance from a fundamental point of view; however, many realistic, nano-technological applications such as NEMS/MEMS are dominated by multi-asperity contacts~\cite{Kim2007a,Mo2009}. In such systems nanowear poses a major limitation for the lifetime, performance, reliability, or the overall usability. Thus, it is imperative to investigate such multi-asperity contacts exhaustively. A further pressing issue regarding contacts at the nanoscale is the calculation of the real contact area. Commonly applied continuum mechanics approaches are hardly applicable at this length scale~\cite{Luan2005}; therefore, various methods have been proposed to calculate the contact area based on atomistic principles, see for example refs~\citenum{Mo2009,Cheng2010,Wolloch2015}.

As addition to experimental techniques and to obtain a more complete picture, classical molecular dynamics (MD) simulations have proven very useful to simulate nanoscale tribological systems~\cite{Szlufarska2008}. Modern computing power allows tracking the time development of fully atomistic systems consisting of several million atoms without the \emph{a priori} assumption of constitutive equations governing the systems' behavior, see refs~\citenum{Goel2015,Brinksmeier2006} and the references therein for some examples. One important aspect of non-equlilbrium MD (NEMD) simulations, i.e., where energy is continuously introduced into the system via external constraints such as imposed forces or velocities, is the thermodynamically correct removal of the generated heat. Several concepts have been proposed to do this in a manner consistent with some thermodynamic ensemble, such as the Nos\'e-Hoover~\cite{Nose1984,Hoover1985}, the Berendsen~\cite{Berendsen1984}, or the Langevin thermostats~\cite{Schneider1978,Dunweg1991}, to name only a few.

The feasibility and quality of a NEMD simulation is strongly affected by the choice of the thermostat as well as its subsequent configuration. Particularly, this includes the decisions on which parts of the system the chosen thermostat acts upon and on the coupling strength between system and thermostat. In the following, three distinct scenarios will be presented. For a nanomachining simulation as considered in this work it would, in principle, make sense to apply the thermostat only to a part of the system some distance away from the tribological interface. Such a setup would minimize the interference with the physics of the processes occurring close to the surface. This implies that there is a large heat sink attached to the base of the explicitly considered near-surface region, namely the bulk. Unfortunately, typical interaction potentials modeling the behavior of metals, such as EAM~\cite{Daw1984} or Finnis-Sinclair~\cite{Finnis1984}, do not explicitly include electrons and are therefore not able to sufficiently reproduce the thermal conductivity of a metal. As the modeled systems grow ever larger, resulting in larger distances between contact zone and heat sink, the grossly overestimated temperature gradients additionally lead to highly unrealistic absolute temperature differences. We will henceforth refer to this scenario as ``base-thermostatted''. 

An alternative approach is to thermostat the entire work piece. Most frequently, the thermostat is strongly coupled to the atoms so that heat is almost instantaneously transported away. While this certainly allows one to maintain good temperature control over the work piece, one may completely neglect the thermal influence of the modeled mechanical process itself by eliminating heat conduction altogether. We will call this scenario ``fully thermostatted''.

When it comes to atomistic simulations of systems featuring considerable heat transfer through metals, many authors do not mention which thermostatting method they use to control the system's temperature. If they do, they seldom disclose how they parametrized the thermostat or justify why they chose the particular parametrization. Without any claim to completeness, we list some examples of otherwise high-quality simulations that are very likely underthermostatted~\cite{Maekawa1995,Jeng2005,Karthikeyan2009,Li2014} due to base-thermostatting or possibly overthermostatted~\cite{Ye2003,Shiari2007,Chamani2016,Chien2016} due to full thermostatting. Out of these examples, it seems that only Shiari \emph{et al.}~\cite{Shiari2007} are aware of some implications of their simplified thermostatting efforts and concede that another group reported that material removal is greatly influenced by the thermal conditions in the shear zone~\cite{Shimada1995}.

Gill~\cite{Gill2010} gives a good overview of NEMD as well as multiscale modeling of heat conduction in solids. However, most of this extensive review focuses on non-metallic materials. The presented methods intended for metals are either quasi-static, such as the two-temperature method mainly used for modeling the laser annealing of voids~\cite{Huang2008,Schafer2002}, while others require a multiscale approach to the problem via dynamic coarse graining~\cite{Liu2007a} or featuring coupling schemes to continuum~\cite{Schall2005}. The latter apply their \emph{ad hoc} technique to an examination of frictional heating during sliding by solving the heat equation and imposing a thermal conductivity.

In this work, we adopt the concept of electron-phonon coupling as laid out in refs~\citenum{Caro1989,Hou2000}. This does not require us to implement a time-consuming multiscale approach, but rather assumes that due to their high mobility within the metal, the electrons can be used as an implicit heat sink permeating the substrate. So far, this setup is identical to what we described above as ``fully thermostatted''. The crucial point of putting the electron-phonon coupling concept into practice is correctly parametrizing the thermostat so that it reflects how the electrons in the metal interact with the lattice vibrations. We will consider the two extreme cases of thermostatting only the substrate base (``base thermostatting'') and thermostatting the entire substrate with a strongly coupled thermostat (``full thermostatting'') and compare these two limiting scenarios to results obtained with a carefully parametrized thermostat that attempts to reflect the electron-phonon coupling in the metallic work piece as closely as possible. Additionally, we check if considerably extending the heat treatment period of the initial substrate configuration significantly changes the results of the abrasion simulations. We discuss how the application of the electron-phonon coupled thermostatting scheme affects some typically evaluated tribological quantities such as friction, material removal/wear, contact area, and surface topography, as well as the microstructural development of the work piece. This constitutes an important step towards making the results of atomistic non-equilibrium simulations of materials undergoing plastic deformation and subjected to high temperature gradients more reliable.


\section{Modeling and Simulation Details}
\label{sect:ModSim} 

\begin{figure*}[hbtp]
\centering
\includegraphics[width=0.99\textwidth]{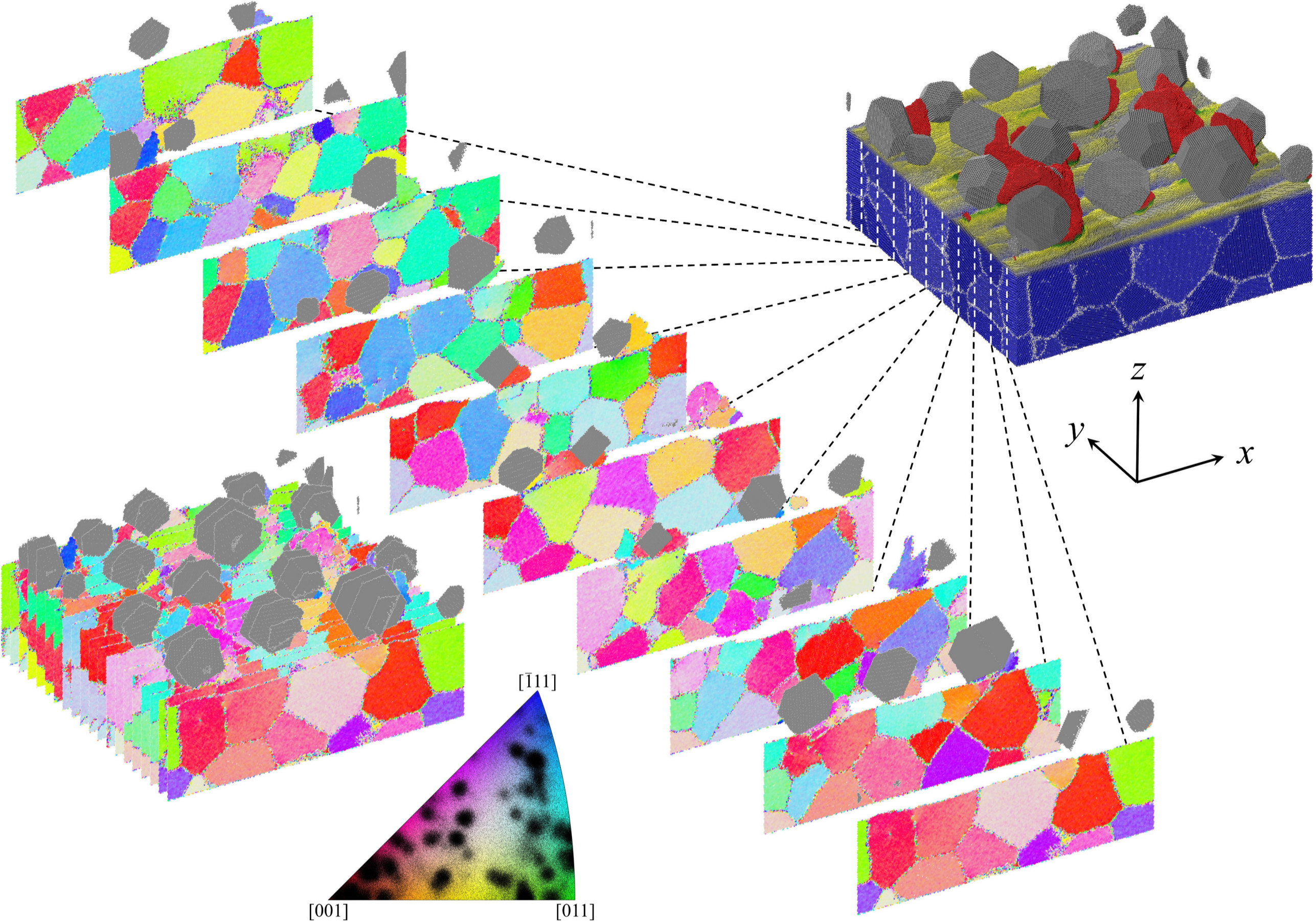}
\caption{\label{fig:EULtomo} The image in the top right is a 3D representation of the entire model during the nanomachining process. The bulk has grain boundary coloring (white on blue), whereas the surface has topographic coloring (blue--yellow--green--red, from low to high) so that the abraded material is shown in red. Abrasives are gray. Machining effectively takes place in $+x$ direction. Computational substrate tomography is performed by decomposing the system into 20 $xz$ slices normal to the $y$ direction, one every 3~nm. In this example, the atoms are colored according to the grain orientation similar to electron backscatter diffraction (inverse pole figure standard, see legend at the bottom with black point clouds representing the individual grains). The tomographs are finally arranged on a 4$\times$5 grid for a better overview, see Fig.~\ref{fig:sliceEul_HT1HT2}.}
\end{figure*}

All our simulations were carried out using the open-source MD code LAMMPS~\cite{Plimpton1995}. The exact procedure of our model construction is described in an earlier work~\cite{Eder2017a}. It consists of a rough 60$\times$60$\times$20 nm$^3$ $\alpha$-Fe polycrystal including randomly oriented abrasives with Gaussian size-distri\-bution. A representative snapshot of the 3D model is shown in the top right corner of Fig.~\ref{fig:EULtomo}, along with some diagonally arranged tomographs colored according to the grain orientation. The mean equivalent diameter of the bcc Fe grains in the originally constructed 3D-periodic 60$\times$60$\times$60 nm$^3$ cube is 12.7~nm, corresponding to an average grain volume of 1080~nm$^3$. Due to the introduction of the surface at $z=20$~nm, out of the 128 remaining grains, the cleft ones no longer have the shape and size of their original Voronoi cells. Therefore, the final overall size distribution of the grains features 40 grains with volumes of less than 200~nm$^3$, a size that was not present in the original size distribution at all, while 20 grains have sizes between 200~nm$^3$ and 400~nm$^3$, compared with only 7 in the original polycrystal. There are thus approximately 75 grains in the initial substrate that almost fully retain their original geometry. The grains are initially randomly oriented, and the surface topography has a fractal dimension of 2.111, an RMS roughness of 0.7~nm, and a lower frequency cut-off producing a typical lateral roughness feature extent of 23~nm. The atoms located in the lower 3~\AA\, of the simulation box are kept rigid to emulate bulk support, and the Fe--Fe interactions within the substrate are governed by a Finnis-Sinclair potential~\cite{Mendelev2003}. 

\begin{figure*}[hbtp]
\centering
\includegraphics[width=0.99\textwidth]{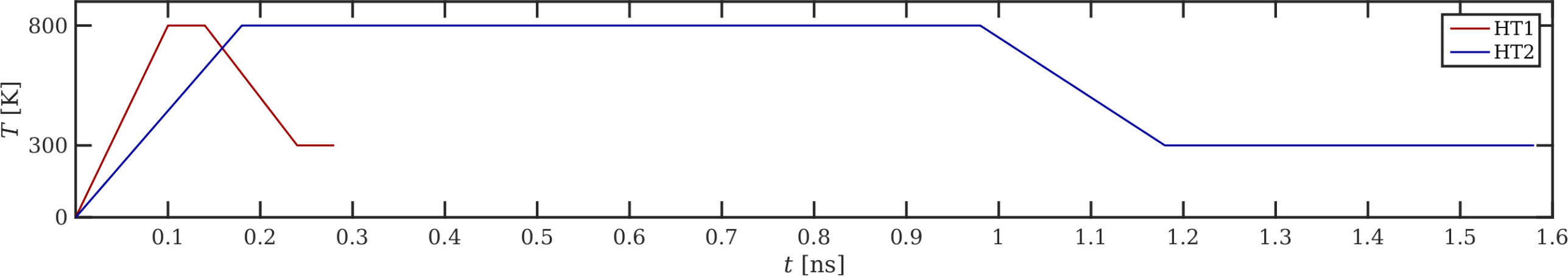}
\caption{\label{fig:HT1_HT2} Temperature curves for the heat treatments ``HT1'' and ``HT2'' leading to the two initial system configurations used in this work.}
\end{figure*}

\begin{figure*}[hbtp]
\centering
\includegraphics[width=0.99\textwidth]{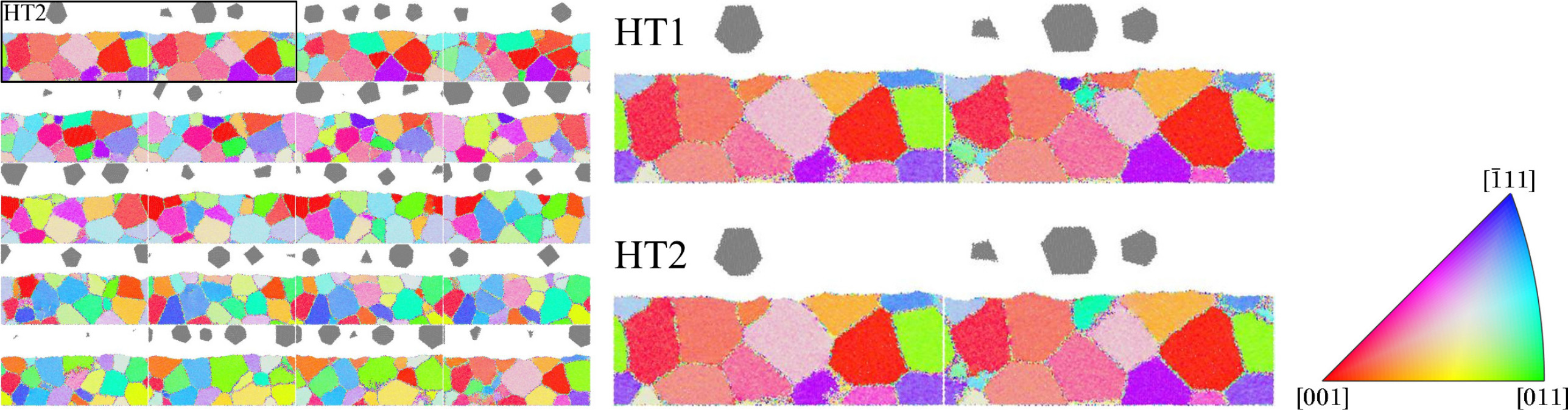}
\caption{\label{fig:sliceEul_HT1HT2} Left: Substrate tomographs of the initial configuration after heat treatment 2 (HT2), sorted from top left to bottom right. Center: Close-up of exemplary slices \#1 and \#2 (marked by box in left panel) for comparison between HT1 and HT2. Abrasives are gray, grains are colored according to orientation, inverse pole figure (IPF) standard, see legend on the right.}
\end{figure*}

Two slightly different initial configurations were produced by subjecting them to two different types of heat treatments, abbreviated ``HT1'' and ``HT2'' henceforth, with the temperature curves shown in Fig.~\ref{fig:HT1_HT2}. In both heat treatments, a Langevin thermostat~\cite{Schneider1978,Dunweg1991} controlled the temperature of the entire substrate (abbreviated ``full'' in some figures) during the ramps, while during the constant temperature intervals only a layer with a height of 0.3~nm at the substrate base was thermostatted (``base''). In this way, the annealing process is efficient while allowing the crystal grains to reorganize without too much thermostat interference. This basic Langevin thermostat was parameterized with a time constant of 0.5~ps, corresponding to strong electron-phonon coupling in metals~\cite{Halte1998}. For a comparison between the substrate microstructures after the two heat treatments, see the selected substrate tomographs in the center of Fig.~\ref{fig:sliceEul_HT1HT2}. These tomographs are colored according to grain orientation as in electron backscatter diffraction, using the inverse pole figure coloring standard. The orientations were calculated using polyhedral template matching~\cite{Larsen2016} as implemented in OVITO~\cite{Stukowski2009}, and the color rendition was carried out using the MTEX toolbox~\cite{Bachmann2010,Nolze2016} for Matlab. The differences between HT1 and HT2 are subtle, even when magnified, and best visible in or around small near-surface grains, see the comparison for slice \#1 shown in the center of Fig.~\ref{fig:sliceEul_HT1HT2}. The comparison for slice \#2 (right side) gives an idea of the largest occurring differences between the two heat treatments, where the small purple and orange grains located in the upper central part of the section have disappeared in favor of neighboring grains after HT2.

The 18 abrasives are modeled as rigid truncated octahedra with diameters ranging from 6--14~nm, randomly rotated and distributed laterally, leading to an effective surface coverage of 37\%. As in ref~\citenum{Eder2015b}, the interactions between abrasive particles and the Fe surface are modeled using a Lennard-Jones potential with the parameters $\varepsilon_\mathrm{LJ}=0.125$~eV and $\sigma_\mathrm{LJ}=0.2203$~nm. These values are similar to refs~\citenum{Maekawa1995,Pei2009} and yield an interaction roughly one order of magnitude weaker than the one for Fe--Fe, which means that there will be some adhesion between the abrasives and substrate as well as the wear particles. The abrasives are pulled across the surface at a sliding velocity in $x$ direction of $v_\mathrm{(slide)}=80$~m/s and an angle of $6.42$° with the $x$-axis, so that they re-enter the simulation box at different $y$ positions every time they pass the periodic box boundaries and therefore never follow exactly in their own grinding marks. The relative abrasive positions are locked throughout the simulations, and abrasive rotation is disabled. Furthermore, the abrasives can change their $z$ position collectively depending on the topography, similar to a grindstone, but not individually. Because of their rigidity, the abrasive particles themselves are not subject to any wear. The normal pressure $\sigma_z$ on the abrasives (defined as the total force in $-z$ direction acting on the abrasive particles divided by the lateral cross-section of the simulation box, 3595.4~nm$^2$) is kept constant at values ranging from 0.1 to 0.9~GPa for a simulation time of 5~ns. During the grinding simulations, the Langevin thermostat acts only in $y$ direction, (nearly) perpendicular to the directions of normal pressure and grinding, so as not to overly interfere with these external constraints.


\section{Thermostat Parametrization}
\label{sect:ThermPar} 

The electronic contribution to the thermal conductivity $\kappa(T)$, which dominates in metallic systems, is not accounted for in the interaction potentials applied in classical MD. We calculated $\kappa(T)$ obtained from the state-of-the-art Finnis-Sinclair potential for Fe used in this work~\cite{Mendelev2003}, which is based only on the phononic contribution, for temperatures ranging from 300 to 900~K using the Green-Kubo formulas, which relate the ensemble average of the auto-correlation of the heat flux $q$ to $\kappa(T)$. When comparing the results to the experimental TPRC data series~\cite{Touloukian1970}, we found that $\kappa(T)$ calculated via MD correlates well with the experimental values, but that the MD values are, on average, 4.5 times lower than the experimental ones. 

We will now describe a ``smart'' thermostatting scheme, based on refs~\citenum{Caro1989,Hou2000}, which attempts to reflect the coupling between electrons and phonons in metals. 
Such an approach is justified as long as the externally imposed velocities lie well below the speed of sound in the work piece, and the mean free path for the electron-phonon interactions is short enough to be accommodated in the work piece along the direction of heat conduction. With our grinding velocity amounting to less than 2\% of the speed of sound in iron, the first condition is easily met. An estimation based on the assumptions in ref~\citenum{Chantrenne2003} yields an electron-phonon mean free path of approximately 2.5~nm, which is roughly one order of magnitude smaller than our work piece thickness.
The Langevin thermostat acting on the bulk of the substrate is parametrized with a coupling constant corresponding to the characteristic time that it takes the lattice vibration energy of the ionic system to be transferred to the electron gas, which, due to its high thermal conductivity, can be considered an implicit heat bath. Based on the Sommerfeld theory of metals, an estimation for this electron-phonon coupling time $\lambda$ can be expressed as~\cite{Hou2000}
\begin{equation}
\label{eq:lambda}
 \lambda = \frac{2 m_e \kappa E_F}{\Theta_D T_0 L n e^2 k_B Z} \ ,
\end{equation}
where $T_0$ is the temperature of the heat bath, $\Theta_D$ the Debye temperature, $L$ the Lorenz number, $n$ the free-electron density, $Z$ the valence, $\kappa$ the thermal conductivity, $E_F$ the Fermi energy, $k_B$ the Boltzmann constant, and $e$ and $m_e$ the electron charge and mass. If we feed this expression with the data for Fe at the expected mean substrate temperature of $T\simeq 310$~K, we obtain a $\lambda$ of about 0.5~ps. 

\begin{figure*}[hbtp]
\centering
\includegraphics[width=0.99\textwidth]{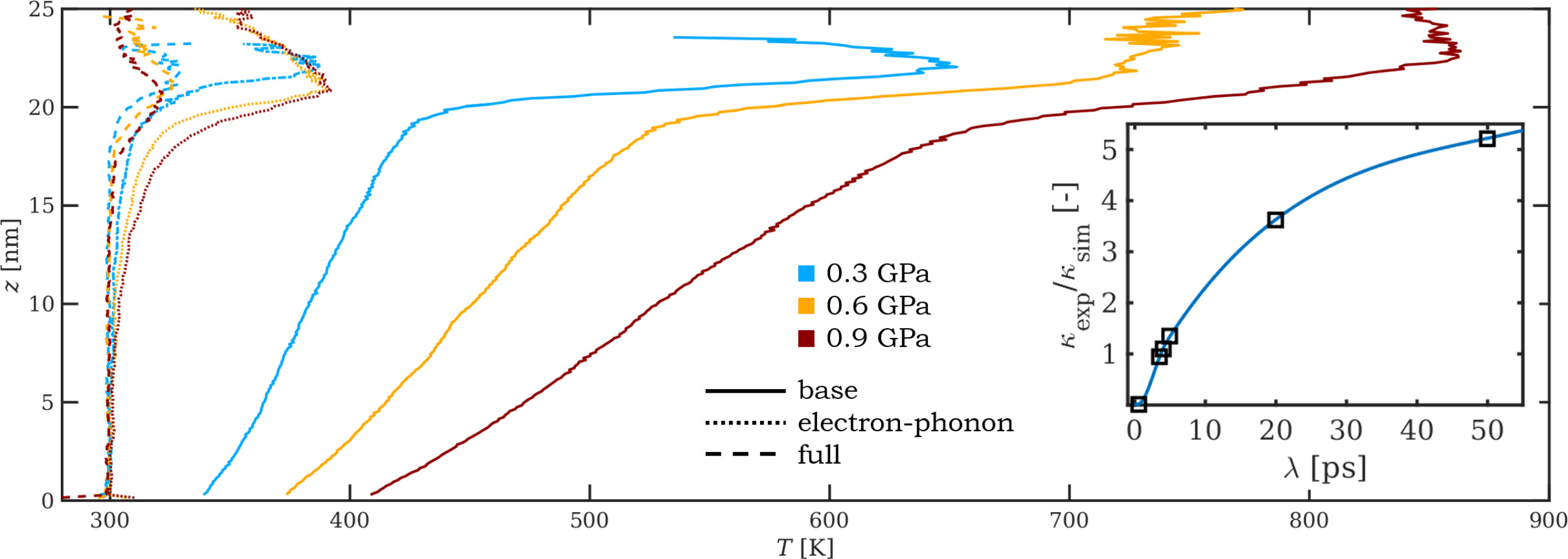}
\caption{\label{fig:q4lambda} Temperature profiles for three different normal pressures and three different thermostatting procedures. The inset shows the dependence of the ratio between the experimental thermal conductivity and the value estimated from the simulations, $\kappa_\mathrm{exp}/\kappa_\mathrm{sim}$ on the electron-phonon coupling time $\lambda$. The black boxes represent values obtained from simulations at 0.9~GPa, and the blue curve is a B-spline interpolant.}
\end{figure*}

On the other hand, we can roughly estimate the thermal conductivity of our substrate by assuming that almost all the frictional energy dissipated in the interface must find its way to the heat bath eventually. We neglect the heat that flows from the sliding interface into the wear particles as well as radiation losses from the surface. The former contribution depends on the load and amounts to several percent of the total, while the latter are not even included in our MD model, but are roughly six orders of magnitude smaller than the friction power at the surface temperatures we encounter in our simulations. We thus set the one-dimensional heat flux to $\Phi_q = \mu \sigma_z  v_\mathrm{(slide)}$, where $\mu$ is the coefficient of friction, $\sigma_z$ is the normal pressure, and $v_\mathrm{(slide)}$ is the sliding speed. While $\sigma_z$ and $v_\mathrm{(slide)}$ are kept constant as external constraints, $\mu$ turns out to maintain a value of 1 throughout the simulations discussed in this work (see Sec.~\ref{sect:TriboQuant} for details), so we may also treat it as a constant. The effective thermal conductivity measured from our grinding simulations can then be obtained via the Fourier law
\begin{equation}
\label{eq:kappa}
 \Phi_q = -\kappa(T) \frac{dT(z)}{dz}\ .
\end{equation}
We can now search for the optimum value of $\lambda$ starting from the value 0.5~ps estimated using Eq.~\eqref{eq:lambda} to parametrize our Langevin thermostat for practical use. This is done by performing an abrasion simulation at constant $\Phi_q$ up to the point where it reaches a sufficiently linear temperature gradient $dT(z)/dz$ within the substrate close to the machining interface and checking how well that gradient corresponds to the experimental thermal conductivity of Fe~\cite{Touloukian1970} at the mean substrate temperature, see Fig.~\ref{fig:q4lambda} for some associated temperature profiles at various loads and values of $\lambda$. We calculated $\kappa(T)$ for several values of $\lambda$ in this way and found that the best reproduction of the experimental thermal conductivity of Fe can be achieved with $\lambda = 3.5$~ps throughout our desired range of normal pressures, c.f. the inset in Fig.~\ref{fig:q4lambda}. Simulations of ion beam mixing and damage production in Fe~\cite{Bjorkas2009} with an implementation of the electron-phonon coupling model of ref~\citenum{Hou2000} also came to the conclusion that the time constant best reflecting the experimental data was several times larger than the one calculated via Eq.~\eqref{eq:lambda}, arriving at values between 1.6 and 2.2~ps. Other work dealing with metal ablation by picosecond laser pulses~\cite{Schafer2002} arrives at an electron-phonon coupling time of 7~ps. We can thus conclude that the estimate for $\lambda$ calculated via Eq.~\eqref{eq:lambda} can only serve as a basis for empirically finding the electron-phonon coupling time that best reflects experimentally observable quantities. Table~\ref{tbl:lambdas} summarizes the coupling times parametrizing the Langevin thermostats acting on the base and the bulk of the substrate to implement the three discussed thermostatting schemes.

An additional benefit of the electron-phonon coupled thermostatting scheme discussed above is that the cooling of the removed matter (e.g., via a coolant fluid) is, at least for all practical purposes, accounted for in this way, i.e., even if the wear debris are already detached from the substrate surface and would therefore no longer be thermally coupled to a base thermostat, they can cool off by themselves. 

\begin{table}
\caption{\label{tbl:lambdas} Overview of the coupling times for the Langevin thermostats acting on the substrate base ($\lambda_\mathrm{base}$) and on the rest of the Fe atoms ($\lambda_\mathrm{bulk}$).}
\begin{center}
\begin{tabular}{r|cc}
thermostatting scheme & $\lambda_\mathrm{base}$ [ps] & $\lambda_\mathrm{bulk}$ [ps]\\
 \hline
base  & 0.5 & ---\\
electron-phonon coupled & 0.5 & 3.5\\
full & 0.5 & 0.5\\
\end{tabular}
\end{center}
\end{table}

It should be noted that the theoretically calculated maximum surface temperature occurring due to dry sliding~\cite{Vick2000,Vick2003,Abdel-Aal2003} can rise considerably higher than the surface temperatures we observe in our simulations. This is because firstly our simulations last only 5~ns, while macroscopic equilibrium temperatures are reached after much longer periods of sliding. Secondly, the Langevin thermostat representing the heat sink of the work piece is placed only 20~nm from the friction interface while being kept at a constant and rather low temperature of 300~K, so that by fixing the allowed temperature gradient in the near-surface region, we also limit the maximum surface temperature.


\section{Results and Discussion}
\label{sect:ResDisc} 

\subsection{Time, load, and temperature dependence of tribological quantities} 
\label{sect:TriboQuant}
\begin{figure*}[hbtp]
\centering
\includegraphics[width=0.90\textwidth]{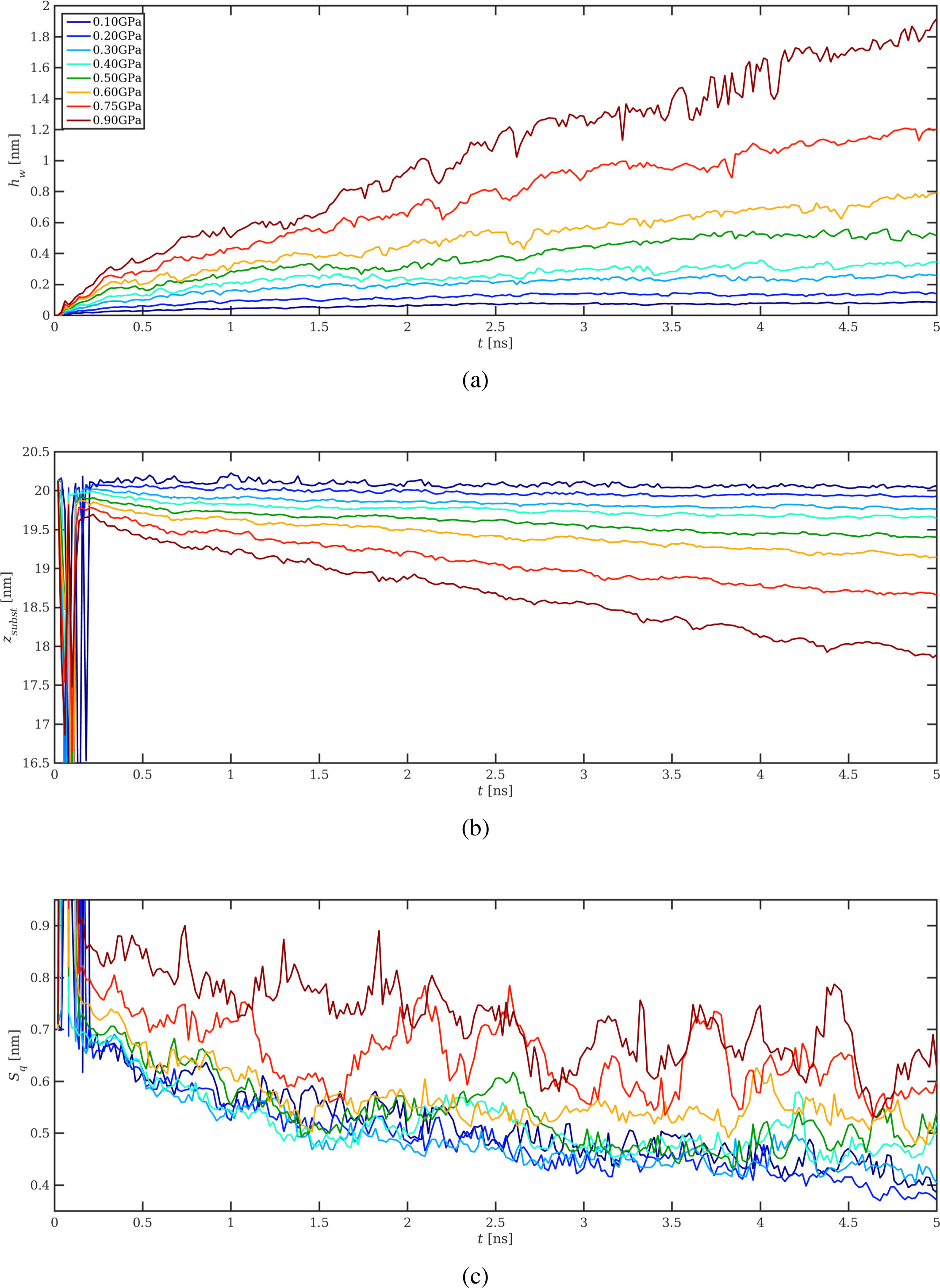}
\caption{\label{fig:Vw_Sq} Wear height $h_w$ (a), arithmetic mean height $z_\mathrm{subst}$ (b), and root-mean-square roughness $S_q$ (c) over time for electron-phonon coupled thermostatting. The rainbow-style coloring reflects the normal pressure (blue is low, red is high).}
\end{figure*}

In Fig.~\ref{fig:Vw_Sq}, we present the time development of the wear depth, the arithmetic mean height, and the RMS roughness of the surface topography for electron-phonon coupled thermostatting. The wear depth is defined as the combined volume of all Fe atoms traveling faster than 90\% of the abrasives' speed, so that they can be considered attached to the grinding particles~\cite{Eder2014a}, divided by the cross-section of the simulation box $A_\mathrm{nom}=3595$~nm$^2$. Compared to our previous work on monocrystalline ferrite substrates and considerably more regular lateral distributions of abrasive particles~\cite{Eder2015b,Eder2015,Eder2015a,Eder2016}, the transition between the running-in and the steady-state regime is much less pronounced, see the kink in most curves in Fig.~\ref{fig:Vw_Sq}~(a) at $t\approx 0.3$~ns. As, by our definition, the substrate consists of all atoms moving in $+x$ direction at less than 10\% of the grinding speed, the substrate topography does not include the shear zone~\cite{Eder2014a}. However, during the initial impact of the abrasives on the substrate and the brief period following it, which is characterized by strong oscillations of the contact forces due to the substrate elasticity, a shear zone is formed that can momentarily be up to two orders of magnitude larger than its equilibrium size and may extend up to 10~nm into the substrate. This explains the large deviations from the general trend during the first 0.2~ns in panels (b) and (c), where the topographic quantities are shown.

The time evolution of the wear depth exhibits a steady increase for the high pressure simulations, which is reflected in a steady decrease of the mean substrate height $z_\mathrm{subst}$. The low pressure simulations hardly show any reduction of $z_\mathrm{subst}$ and a very small increase of the wear depth $h_w$, restricted to the first 0.5~ns. This suggests that up to 0.3~GPa the system does not wear significantly, but that the reduction of the surface roughness, as depicted in Fig.~\ref{fig:Vw_Sq}~(c), is merely a result of material transfer from the peaks into the valleys, i.e., abrasion of the asperity tips and subsequent re-crystallization onto the substrate. The wear particles are formed at the very beginning for low normal pressures, and although individual particles may change their sizes over time, the combined volume of all the wear particles remains fairly constant for the rest of the grinding simulation. For the high pressure cases of 0.9~GPa and 0.75~GPa, the time evolution of $S_q$ is unstable, and the final roughness approximately equals the initial value.

\begin{figure*}[hbtp]
\centering
\includegraphics[width=0.65\textwidth]{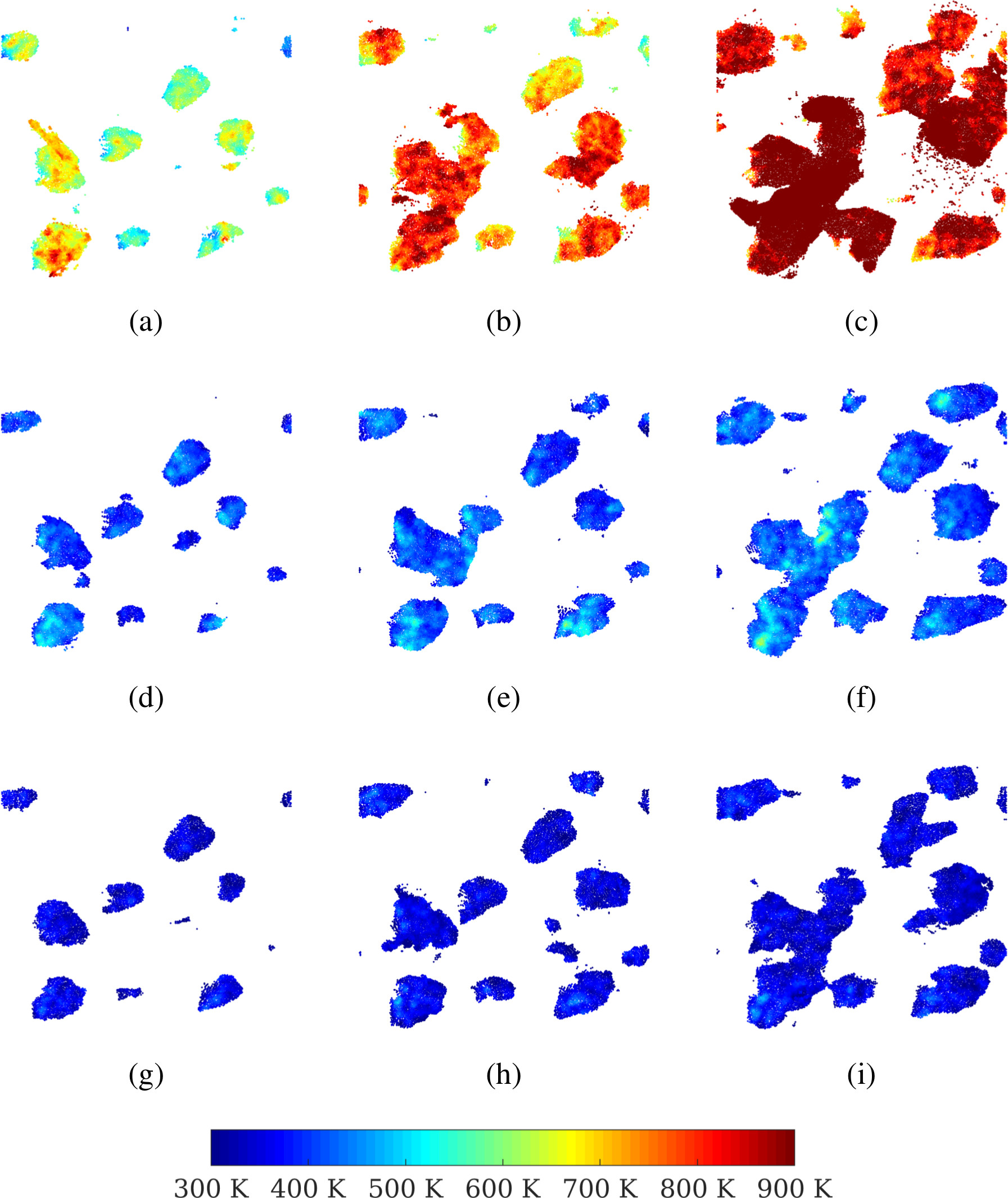}
\caption{\label{fig:AcTc} Exemplary lateral distribution of the contact area ($xy$ projection = top view) and the thermal distribution within the contact area as a function of normal pressure and thermostatting procedure at $t=5$~ns. Top: base-thermostatted substrates, center: electron-phonon coupling thermostatted substrates, bottom: fully thermostatted substrates. Left: $\sigma_z = 0.3$~GPa, center: $\sigma_z = 0.6$~GPa, right: $\sigma_z = 0.9$~GPa.}
\end{figure*}

In Fig.~\ref{fig:AcTc}, we show exemplary $xy$ projections of the real atomic contact area, colored according to contact temperature after 5~ns of grinding. As explained in detail in refs~\citenum{Eder2014a,Eder2014}, this contact area is obtained by establishing which substrate atoms are in contact with the abrasives or the associated wear particles via a distance criterion, and then multiplying the number of these contact atoms with a constant per-atom contact area. The contact temperature for each contact atom is calculated via the kinetic energy averaged over a spherical control volume with a radius of 1~nm after subtracting the drift velocity of this control volume, cf. ref~\citenum{Eder2017a}. The three examples shown in Fig.~\ref{fig:AcTc}, from left to right, are for normal pressures of 0.3, 0.6, and 0.9~GPa, respectively. The top row shows results for base-thermostatted substrates, the center one for electron-phonon coupled thermostatting, and the bottom row fully thermostatted substrates. The contact area increases with load independently of the chosen thermostatting procedure. In detail, the shape and size of individual patches of the contact area are different depending on the chosen thermostatting. However, its more general characteristics remain unchanged since these are primarily determined by the asperities of the substrate as well as the size and distribution of the abrasive particles. Clearly, the contact temperature dependence on the normal pressure is considerable when applying only a base thermostat. For the electron-phonon coupling and the fully thermostatted substrate, increasing the normal pressure does not influence the contact temperature much. The main difference between these two variants is the temperature of the hottest spots within the contact area, which is about 100~K higher if electron-phonon coupling is applied. Additionally, the local temperature distribution within the contact area patches is more inhomogeneous for electron-phonon coupling.

\begin{figure*}[hbtp]
\centering
\includegraphics[width=0.99\textwidth]{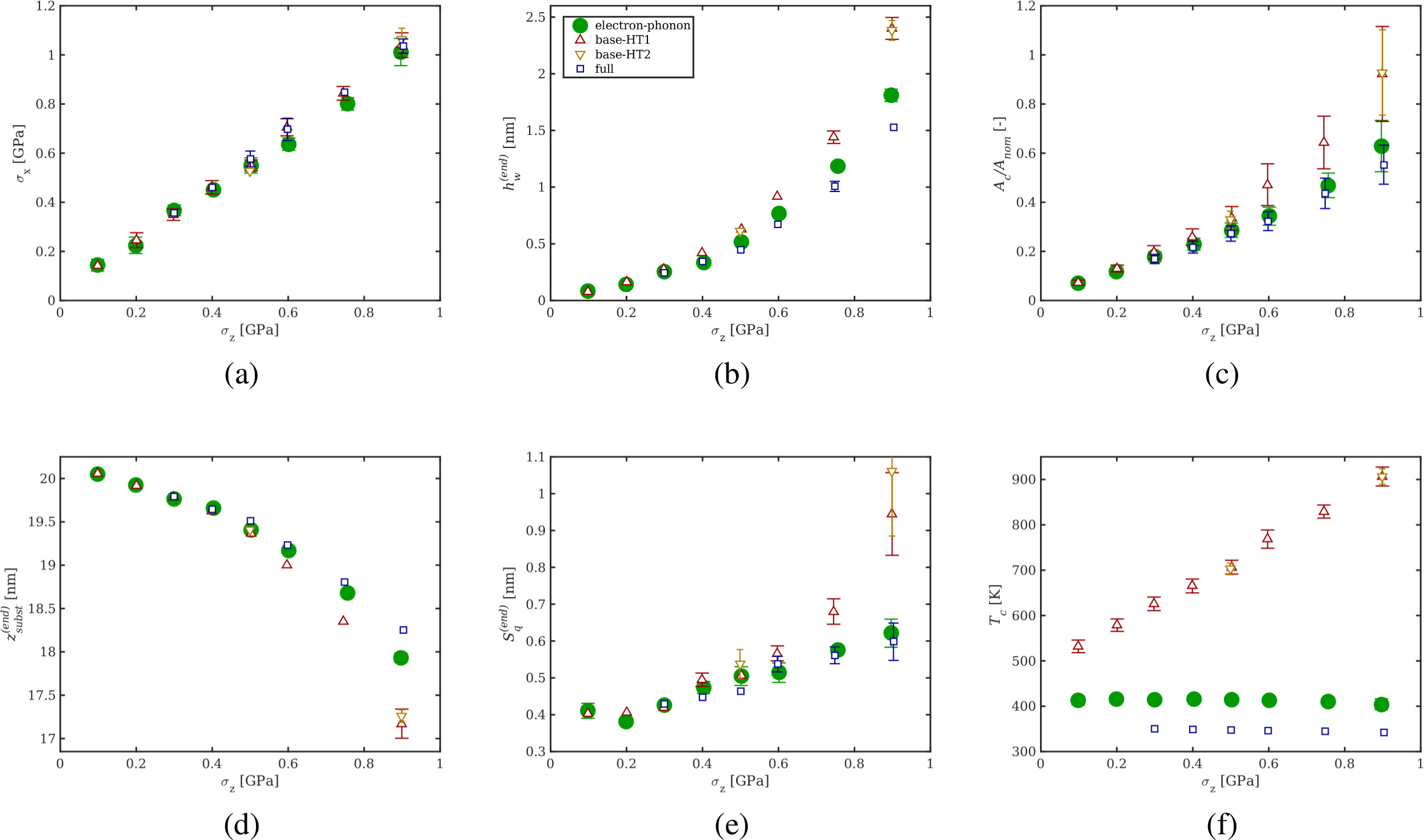}
\caption{\label{fig:allOverSigZ} Mean shear stress $\sigma_x$ (a), final wear depth $h_w$ (b), mean normalized real contact area $A_c/A_\mathrm{nom}$  with $A_\mathrm{nom}=3595$~nm$^2$ (c), final arithmetic mean height $z_\mathrm{subst}$ (d), final root-mean-square roughness $S_q$ (e), and mean contact temperature $T_c$ (f) over normal pressure $\sigma_z$. Full green circles denote results obtained for electron-phonon coupled thermostatting, red upwards triangles for base thermostatting with heat treatment 1 (short), yellow downwards triangles for base thermostatting with heat treatment 2 (long), and blue squares for full thermostatting.}
\end{figure*}

It seems that the choice of the heat treatment procedure has little influence on the normal pressure dependence of some tribological, topographical, and thermodynamic quantities shown in Fig.~\ref{fig:allOverSigZ}. The only noticeable difference occurs for the final RMS roughness at the highest load, see Fig.~\ref{fig:allOverSigZ}~(e), but here the error bars are also very large and overlap to such extent that it must be questioned if the difference between HT1 and HT2 is statistically relevant. 

It is more striking that the pressure dependence of the shear stress $\sigma_x$ does not appear to be influenced by the thermostatting procedure at all. This is in agreement with other literature~\cite{Khan2011}, where it is reported that the substrate temperature, varied from 0~K to 500~K, has little effect on what is there called the ``scratching force''. We therefore carried out additional calculations at a considerably higher normal pressure of $\sigma_z=2.5$~GPa with base- and electron-phonon coupled thermostatting. They revealed that under these severe conditions a difference between the obtained shear stresses emerges. While the electron-phonon coupled results remain on the straight line suggested in Fig.~\ref{fig:allOverSigZ}~(a), the results for the base thermostat lie approximately 25\% lower. This might be explained by a pronounced softening effect of the polycrystal due to the high temperatures in the surface zone, where the base-thermostatted substrate ground at $\sigma_z=2.5$~GPa reaches $z$-averaged temperatures of 1160~K, while the $T$-maxima in the contact spots even rise up to 1350~K. Compared to the much lower surface temperatures of only 450~K for the electron-phonon coupled substrate, the hotter and thus softer base-thermostatted substrate opposes much less resistance to shear, equivalent to reduced work hardening of macroscopic samples at high temperatures.

As expected, the wear depth in Fig.~\ref{fig:allOverSigZ}~(b) and the arithmetic mean height in Fig.~\ref{fig:allOverSigZ}~(d) exhibit opposing trends, both showing almost identical behavior for all thermostatting procedures up to 0.4~GPa and a splitting up into three distinct branches for greater normal pressures. The fully thermostatted substrates are worn less than the electron-phonon coupled ones, which in turn wear less than the base-thermostatted ones. The final RMS roughness of the surface in Fig.~\ref{fig:allOverSigZ}~(e) already shows some differences at a normal pressure of 0.4~GPa, but this difference remains small up to 0.6~GPa. At the highest normal pressure, however, the ``best'' achievable roughness of the base-thermostatted substrates is almost double that of the ones obtained with the other two thermostatting procedures. Note that the initial RMS roughness is 0.7~nm, which means that grinding at normal pressures of 0.75~GPa or more with a base-thermostatted substrate does not produce a smoother surface. Whereas the pressure dependence of the final roughness is almost linear for full and electron-phonon coupled thermostatting, which is in good agreement with ref~\citenum{Eder2015b} (where the substrates were so thin that the thermostatting procedure would not have had much influence on the result), the base-thermostatted systems show a strongly superlinear pressure-dependence. This might be attributed to the following: in the base-thermostatted systems, heat transport from the chips towards the heat sink is only possible via phononic heat conduction through the contact regions with the substrate. The chips can therefore reach temperatures close to 1000~K, making them soft. This, in turn, leads to their occasional slumping over to the sides of the abrasives, leaving high scratch ridges in the likewise softened near-surface substrate.
A linear increase of the RMS roughness with the normal pressure implies that the substrate compliance does not change within the substrate volume that interacts with the abrasives, even at higher loads. Only for the base-thermostatted substrate the temperature gradient is steep, and Fig.~\ref{fig:allOverSigZ}~(e) shows that this temperature increase is only relevant above 0.6~GPa, as there is not enough friction energy to generate sufficiently high temperatures at lower pressures.

Figure~\ref{fig:allOverSigZ}~(c) shows the normalized contact area $A_c/A_\mathrm{nom}$ over the normal pressure $\sigma_z$, where $A_\mathrm{nom}=3595$~nm$^2$ is the lateral cross-section of the simulation box. $A_c/A_\mathrm{nom}$ can be considered the degree to which the two counterbodies are in contact. Since this contact is not necessarily flat (in contrast to $A_\mathrm{nom}$), the dimensionless ratio can in principle exceed 1. For the fully thermostatted systems, the normal pressure dependence is basically linear, while for the base-thermostatted systems it exhibits a distinct kink around $\sigma_z\approx 0.5$~GPa, separating the normal pressure dependence into two regimes of smaller and larger slope for lower and higher normal pressures, respectively. For the systems with electron-phonon coupled thermostatting, the data lie in between, but markedly closer to those obtained for the fully thermostatted systems, with an apparent change in slope between $\sigma_z\approx 0.6-0.75$~GPa.

As can be seen in Fig.~\ref{fig:allOverSigZ}~(f), it becomes evident that in the base-thermostatted substrates, the average contact temperature increases linearly with $\sigma_z$ independent of the heat treatment procedure, with a temperature increase of $\approx 45$~K per 100~MPa. For full and electron-phonon coupled thermostatting the contact temperature remains at a constant 350~K and 410~K, respectively. The temperature difference between the heat bath and the contact zone may therefore only be a function of the abrasive particle geometry, the sliding velocity, and the thermostat coupling constant.

\begin{figure*}[hbtp]
\centering
\includegraphics[width=0.85\textwidth]{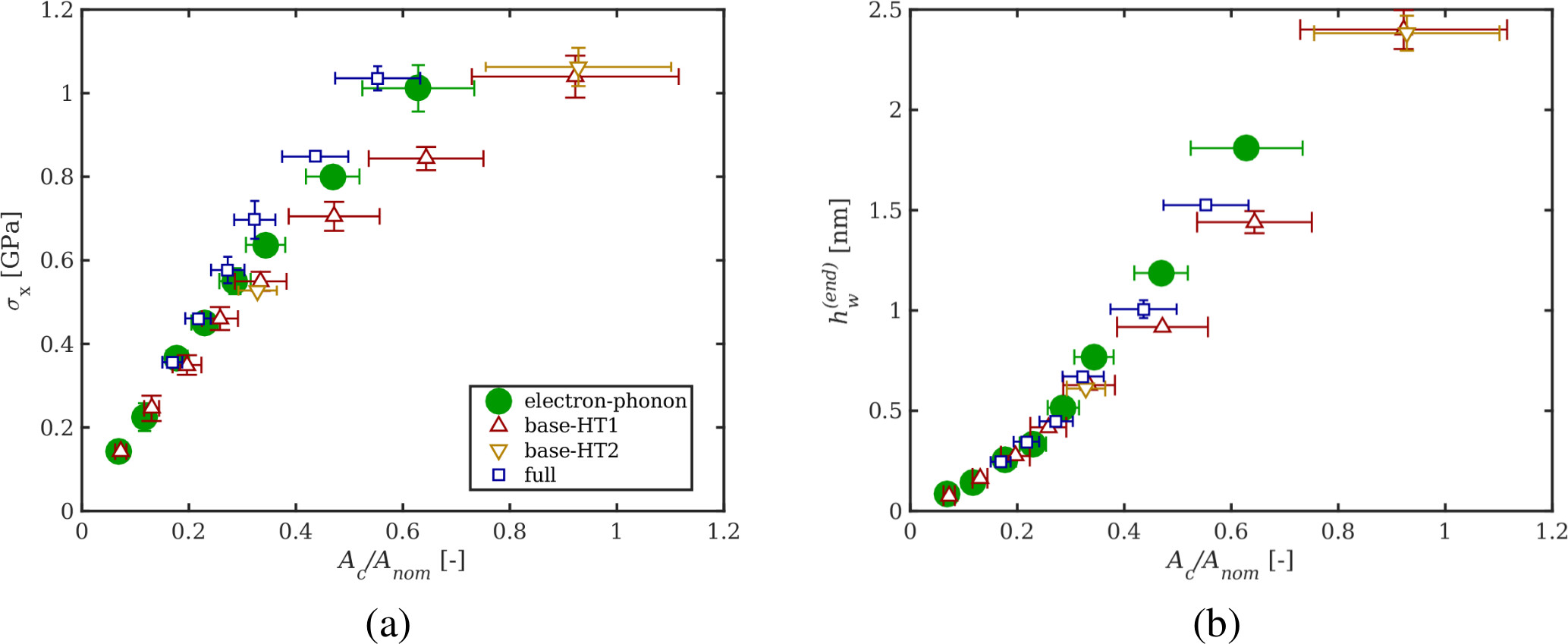} 
\caption{\label{fig:otherDep} Shear stress $\sigma_x$ and final wear depth $h_w^\mathrm{(end)}$ over the normalized contact area $A_c/A_\mathrm{nom}$ with $A_\mathrm{nom}=3595$~nm$^2$. }
\end{figure*}

In Fig.~\ref{fig:otherDep} we analyze how the shear stress $\sigma_x$ and the final wear depth $h_w^\mathrm{(end)}$ vary with the normalized contact area $A_c/A_\mathrm{nom}$. In panel (a), the slope corresponds to the effective shear strength $\tau$ of the system. For the fully thermostatted substrates, this quantity is almost a constant, while for the other two thermostatting procedures we observe some more pronounced flattening of the curve, which reflects the onset of thermal substrate softening. Panel (b) shows that the final wear depth can essentially be considered a superlinear function of the contact area, almost independent of the contact temperature and therefore also the thermostatting procedure. We can therefore formulate that the contact area increases due to thermal softening of the substrate, while the separation of atoms from the substrate is eased at higher temperatures. As this correlation of wear depth and contact area depends on the local temperature of the near-surface region immediately beneath the contact spot, the same superlinear trend is observed irrespective of any thermostatting procedure.

\subsection{Thermostat influence on microstructural development}

\begin{figure*}[hbtp]
\centering
\includegraphics[width=0.85\textwidth]{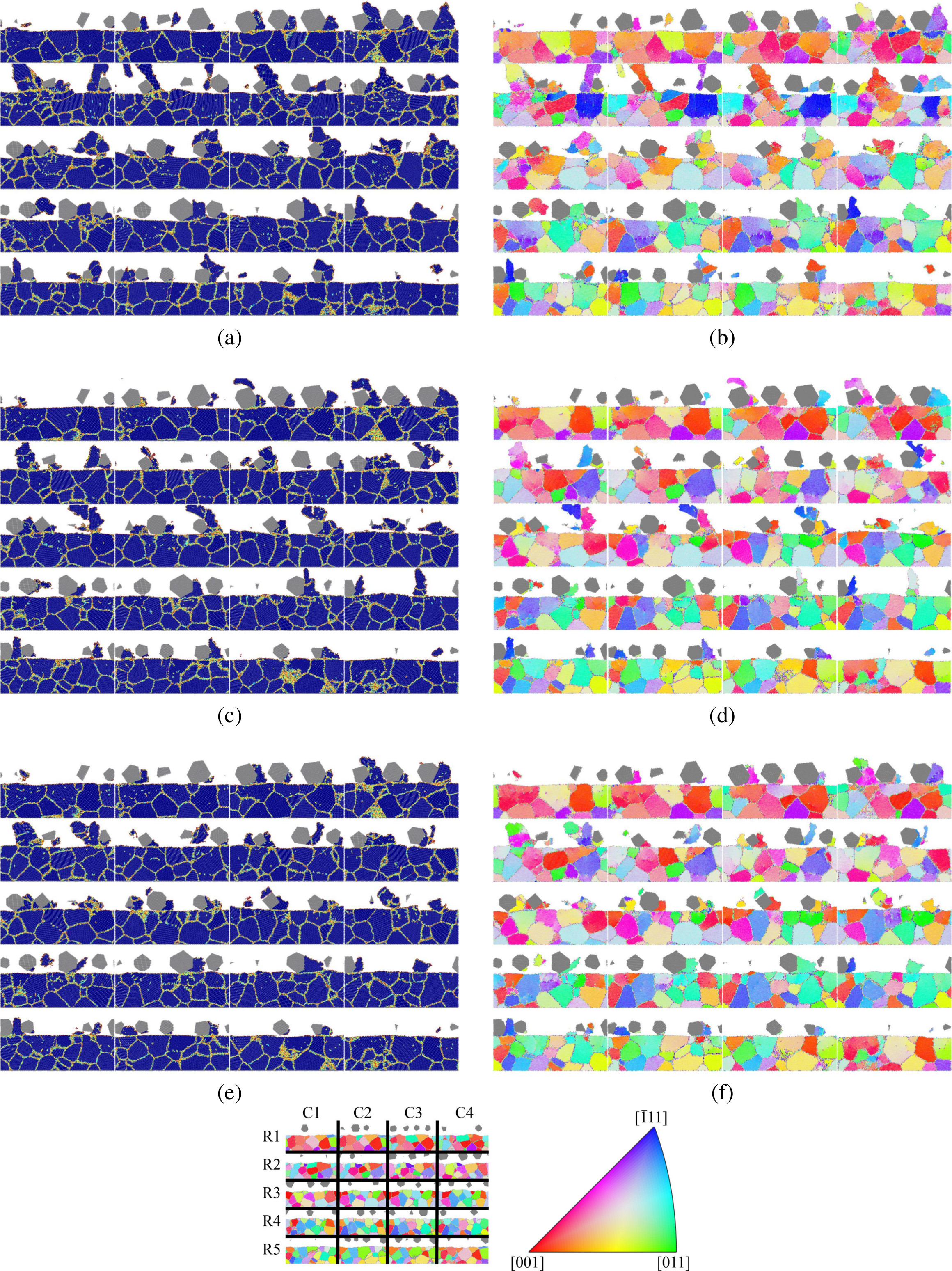}
\caption{\label{fig:slices0.9GPa_CS_IPF} Substrate tomographs after 5~ns of grinding at 0.9~GPa (see legend at bottom left for shorthand references in the text). Left: coloring according to centro-symmetry (CS) parameter (perfect lattice = dark blue, lattice defects = turquoise, grain boundaries = yellow/orange, surface = red), abrasives are gray. Right: grains colored according to orientation (inverse pole figure (IPF) standard, see legend), abrasives are gray. (a,b): base-thermostatted substrate, (c,d): electron-phonon coupled thermostatting, (e,f): fully thermostatted substrate.}
\end{figure*}

\begin{figure*}[hbtp]
\centering
\includegraphics[width=0.85\textwidth]{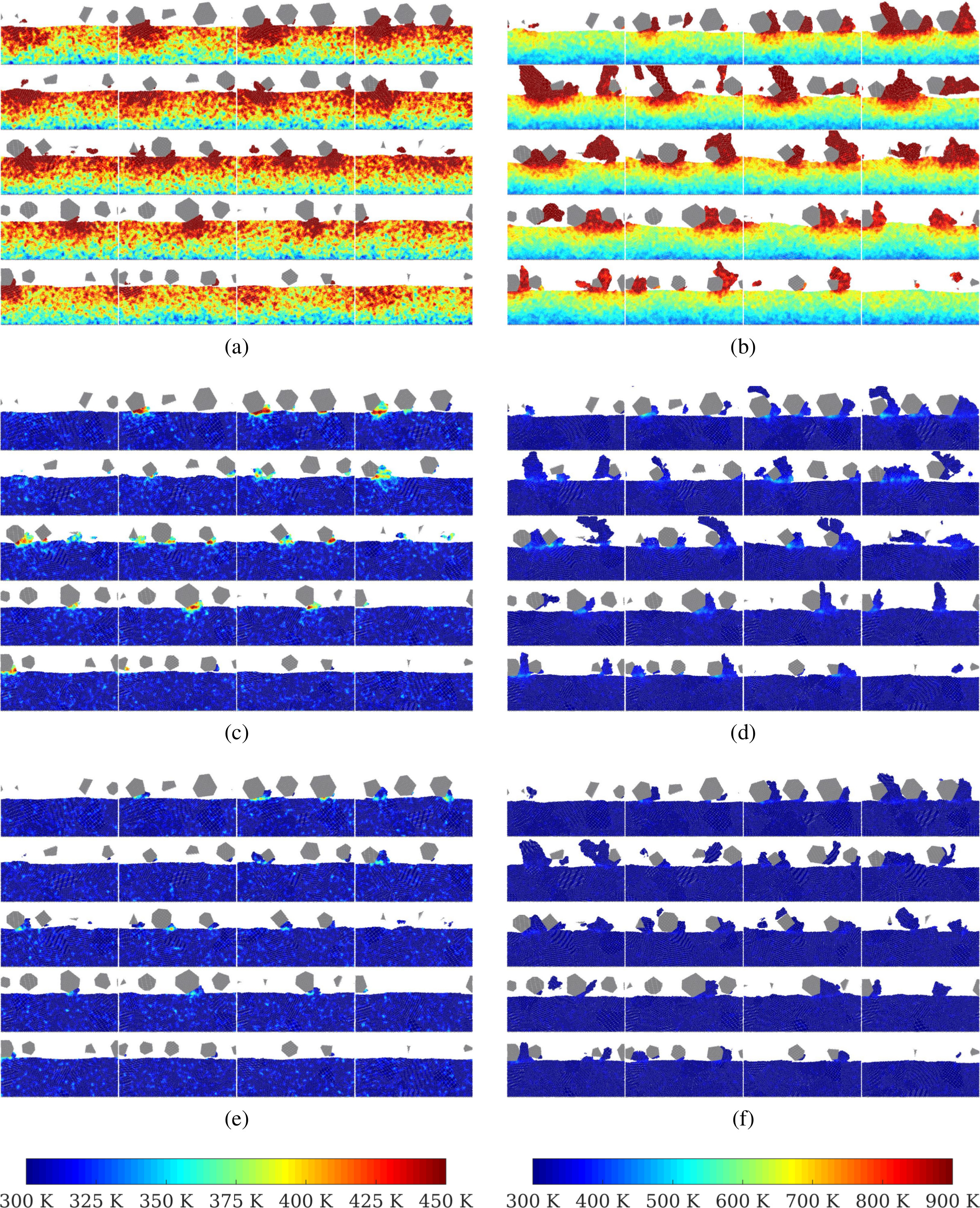}
\caption{\label{fig:slices0.9GPa_T} Substrate tomographs colored according to temperature (see colorbars below, abrasives are gray) after 5~ns of grinding at 0.3~GPa (left) and 0.9~GPa (right). (a,b): base-thermostatted substrate, (c,d): electron-phonon coupled thermostatting, (e,f): fully thermostatted substrate.}
\end{figure*}

The grinding process changes the microstructure as well as the defect structure regardless of thermostatting procedure or normal pressure. These changes are more pronounced at higher normal pressures, and their extent depends critically on the thermostatting procedure. We therefore chose the abrasion simulation carried out at the highest normal pressure (0.9~GPa) to best illustrate the possibly drastic differences between the three thermostatting variants, see Fig.~\ref{fig:slices0.9GPa_CS_IPF}. For an explanation on how the tomographic slices correspond to the 3D model, recall Fig.~\ref{fig:EULtomo}. The analysis scheme using computational substrate tomography is explained in detail in ref~\citenum{Eder2017a}. In the following, individual panels of the substrate tomographs in Figs.~\ref{fig:sliceEul_HT1HT2}, \ref{fig:slices0.9GPa_CS_IPF}, and \ref{fig:slices0.9GPa_T} will be referred to by their row number $m$ and their column number $n$, abbreviated as R$m$/C$n$, cf. the bottom left legend in Fig.~\ref{fig:slices0.9GPa_CS_IPF}. The defect structure evolving due to dislocations as well as any changes of the grain boundaries can be observed via the centro-symmetry (CS) parameter~\cite{Kelchner1998} panels in Fig.~\ref{fig:slices0.9GPa_CS_IPF}~(a,c,e). The differentiation between grain growth and recrystallization or lattice rotation is possible via the orientation panels in Fig.~\ref{fig:slices0.9GPa_CS_IPF}~(b,d,f) with grain coloring according to inverse pole figure (IPF) standard. For an interpretation of the microstructural changes as a result of thermally activated processes, we refer to the temperature plots resulting from low (0.3~GPa) and high pressure simulations (0.9~GPa) in Fig.~\ref{fig:slices0.9GPa_T}. In general, we observe grain growth, abrasion of entire grains, and full or partial rotation of the lattice structure. Part of the friction energy is stored in dislocations or can lead to the formation of new grain boundaries.

The base-thermostatted simulation results in the most dramatic changes to the initial polycrystalline structure, compare Fig.~\ref{fig:sliceEul_HT1HT2}~(a) with Fig.~\ref{fig:slices0.9GPa_CS_IPF}~(b). We observe massive grain growth, and only grain boundaries pinned to the rigid base via the boundary conditions do not experience any changes. The grain structure at the surface is strongly disturbed. As the wear particles cannot cool down by themselves in the base-thermostatted simulations, their temperatures rise beyond 900~K, see Fig.~\ref{fig:slices0.9GPa_T}~(b). Although the surface can cool down somewhat between two passing abrasives, the temperature in the near-surface zone remains high in the base-thermostatted simulation, never dropping below 600~K. The grain boundary mobility in this zone is therefore much higher than in the differently thermostatted simulations. Due to the high temperatures, the zone featuring modified grain orientations extends to depths of 10--15~nm. The latter can be visualized using IPF standard coloring, which reveals massive changes between initial and final grain orientations and a slight tendency towards a (111) orientation, though the orientation is too weak to refer to it as a preferred texture evolving during grinding. 

After grinding at 0.9~GPa, the final near-surface grain size of the electron-phonon coupled substrate is smaller than that of the base-thermostatted substrate. Only beneath the wear particles, the base-thermostatted substrate forms numerous tiny grains down to half the substrate thickness, whereas electron-phonon coupling results in smaller wear particles and hence a reduced grain-refined area, compare R2/C1 in Figs.~\ref{fig:slices0.9GPa_CS_IPF}~(a,c). 

The IPF orientation analysis features pronounced color shading within the larger grains, which indicates elastic deformations within the grains that are also referred to as residual stresses of type II, or microstresses~\cite{Withers2001}. The fact that these elastic deformations are stable in more grains than in the base-thermostatted substrate can be explained by the much lower temperatures in the electron-phonon coupled substrate. Besides, the surface can cool down to almost 300~K between two passing abrasives, and the wear particles themselves are cool and do not heat up the surface as in the base-thermostatted process. At the end of the simulation there is no tendency towards a preferred (111) orientation for the electron-phonon coupled substrate.

We observe grain growth compared to the initial microstructure for the fully thermostatted simulation as well. The CS parameter analysis of the fully thermostatted simulation in Fig.~\ref{fig:slices0.9GPa_CS_IPF}~(e) shows that small grains have been abraded and grain growth has taken place, but only in the first nm beneath the surface. As the temperature of the wear particles and the contact zone only reaches 350~K, any temperature rise due to frictional heat is roughly limited to the upper 3~nm of the substrate, recall Fig.~\ref{fig:q4lambda}. Since grain growth is primarily a thermally activated process, the grain boundary mobility is only high enough in this limited zone of the first nanometers beneath the surface (located at $z\approx 20$~nm). Consequently, the grain sizes at the surface of the fully thermostatted substrate are smaller than in the electron-phonon coupled substrate and much smaller than in the base-thermostatted substrate after 5~ns of grinding, see Fig.~\ref{fig:slices0.9GPa_CS_IPF}~(e,f). The grain orientation analysis in panel (f) reveals that grains grow at the expense of small grains at the surface, and that larger grains located in the middle of the substrate are stable with respect to their orientation and only grow towards the surface. 

The fully and the base-thermostatted scenarios reflect the two extremes of microstructural changes. As the evolution of the microstructure is primarily driven by the evolving temperature (when comparing equal pressure processes), thermostatting strongly changes the temperature levels and gradients at different depths of the substrate. Electron-phonon coupled thermostatting produces temperatures between the two bounds of the fully and base-thermostatted variants, compare Fig.~\ref{fig:slices0.9GPa_T} (c,d) with (a,b) or (e,f). Electron-phonon coupling leads to higher maximum temperature levels in the contact than full thermostatting, but also to elevated temperature fields confined to substrate regions directly beneath the contact area when compared to base thermostatting, cf. Fig.~\ref{fig:q4lambda}. Furthermore, the maximum temperature gradients $dT/dz$ (occurring within the shear zone around $z\approx 20$~nm) are considerably steeper for base thermostatting than for the two other thermostatting modes.

The main microstructural feature that changes during grinding in all simulations is the grain size. Grain growth occurs in part by grain boundary movement and in part by lattice rotation. Occasionally, tiny new grains form in both non-base-thermostatted variants, but they are not sufficiently stable to withstand a passing abrasive.

For the different thermostatting procedures, the resulting temperature gradient within the substrate and the shear zone determines the reaction of the microstructure at the respective normal pressure. The external mechanical loads are fairly equal if the pressure remains fixed. For a comparison of the evolving temperature gradients at low normal pressure (0.3~GPa) for base, electron-phonon coupling, and full thermostatting, see Fig.~\ref{fig:slices0.9GPa_T}~(a,c,e). As the $T$-gradient in the shear zone is very steep for base thermostatting, even this low normal pressure machining process changes the microstructure markedly. The degree to which the microstructure is modified is thus directly defined by the level and extent of the three-dimensional temperature field. 


\section{Conclusion and Outlook}

In this work we have analyzed the sliding interface between a nanocrystalline ferritic work piece and hard, abrasive particles during a nanomachining process simulated with molecular dynamics simulations. We have shown that choosing a thermostatting procedure that can mimic the electronic contribution to heat conductivity is crucial when large amounts of energy are introduced to metallic systems in a highly localized fashion. In our application, the resulting near-surface microstructural development, surface topography, and quantities such as the wear depth depend strongly on the chosen thermostatting procedure for normal pressures exceeding 0.4~GPa, while the total shear stress is almost unaffected up to a normal pressure of approximately 1~GPa. The correlation between plastification of the substrate and varying normal pressures as well as the evolving dislocation structure will be discussed in detail in a forthcoming publication by some of the authors.

As beneficial as the thermostatting approach described in this work is for simulating surface finishing and wear processes, there are several aspects of it that could be improved further. As the target temperature of the thermostat is constant throughout the simulations, it is currently impossible to reach the practically obtained equilibrium temperatures for dry friction with simulation times restricted to several nanoseconds. A simple workaround could be to set the thermostat temperature close to the theoretically predicted value at that depth in the work piece, but a proper treatment would have to include a coupling of the base thermostat to a far-field solution of the temperature depth profile. This could be obtained using a one-dimensional Green's function approach~\cite{Vick2000, Carslaw1959}, assuming an averaged heat flux at the sliding interface and a semi-infinite work piece, thus placing the bulk temperature boundary condition ``far away'' from the heat source. Properly done, this would allow the system to reach a realistic surface temperature. A second aspect that is not trivial to resolve is that the Langevin thermostat somewhat unrealistically damps all modes of heat conduction equally, which might be improved by employing a more sophisticated thermostat. Finally, since nanoscopic heat conduction depends on the crystal system and orientation, it might be an oversimplification to assume a constant macroscopic heat conductivity for a 20~nm thick surface layer, even if it consists of initially randomly oriented grains. A proper treatment of this issue would include a real-time grain orientation analysis and subsequent adaptive thermostat parametrization, so any benefit gained from such an approach would have to be carefully balanced with the considerable additional computational cost.

Although the thermal and microstructural trends in our results match intuition, the highly desirable experimental validation is unlikely to be exhaustive. All of the thermostatting scenarios in our work except the electron-phonon coupled ones are difficult to reproduce in an experiment, as it is not trivial to suppress electronic heat conduction~\cite{Yao2017,Lukas2012}. While one could simply conduct nanomachining trials at higher temperatures, any influence of excessive thermal gradients would still be lost. That said, it may be possible to verify the time-development of the near-surface structural changes occurring in a nanocrystalline sample due to abrasive nanomachining. One aspiring suggestion would be the design of a tribo-experiment to be analyzed \emph{in situ} using synchrotron X-ray diffraction, similar to what has been done to study the growth of bulk grains in deformed metals~\cite{Schmidt2004}. Before going to such an effort, it may however be sensible to reconsider the choice of work piece material, e.g., an fcc structure with a low tendency for oxidation while still being technologically relevant.

There exists a second path for, at least qualitative, validation of our simulation results. Aided by ever-increasing computational power, we have recently succeeded in simulating the tribological response of a considerably larger work piece consisting of approximately 25 million atoms. Due to the initial grain size diameters of 30~nm, tribological loading actually leads to dislocation pile-up and near-surface grain refinement, both of which are observed in macroscopic experiments. This bodes well for the future, as it means that phenomena relevant to applied interfaces might be reproduced at a slightly smaller scale, thus uncovering the underlying mechanisms. Ultimately, MD simulation may well evolve into a crucial component of engineering design tools for manufacturing, friction, or wear applications.


\begin{acknowledgement}
This work was partly funded by the Austrian COMET-Program (Project K2, XTribology, no. 849109) and carried out at the ``Excellence Centre of Tribology''. SJE, UC-B, and DB acknowledge the support of the Province of Nieder\"osterreich (Project ``SaPPS'', WST3-T-8/028-2014). GF acknowledges the financial support from the German Research Foundation (DFG) via SFB 986 ``M$^3$'', project A4. The authors wish to thank M. Moseler, M. M\"user, J. Belak, A. Vernes, and G. Vorlaufer for helpful discussions about thermal conductivity, electron-phonon coupling, and Green's functions.
\end{acknowledgement}

\providecommand*\mcitethebibliography{\thebibliography}
\csname @ifundefined\endcsname{endmcitethebibliography}
  {\let\endmcitethebibliography\endthebibliography}{}

\end{document}